\documentclass[conference]{IEEEtran}
\IEEEoverridecommandlockouts

\usepackage{cite}
\usepackage{amsmath,amssymb,amsfonts}
\usepackage{algorithmic}
\usepackage[draft]{graphicx}
\usepackage{textcomp}
\usepackage{xcolor}
\usepackage{blindtext}
\usepackage{booktabs}
\usepackage{hyperref}
\usepackage{multirow}

\def\BibTeX{{\rm B\kern-.05em{\sc i\kern-.025em b}\kern-.08em
    T\kern-.1667em\lower.7ex\hbox{E}\kern-.125emX}}

\newcommand{\githuburl}{\texttt{https://github.com/HPAI-BSC/RuC}}
\newcommand{\huggingfaceurl}{\texttt{https://huggingface.co/datasets/HPAI-BSC/RuC-datasets}}
    
\begin{document}

\title{RuC: HDL-Agnostic Rule Completion Benchmark Generation
}

\author{
\IEEEauthorblockN{Arnau Ayguadé Domingo\textsuperscript{1}, Miquel Alberti-Binimelis\textsuperscript{1}, Cristian Gutierrez-Gomez\textsuperscript{1}, Emanuele Parisi\textsuperscript{1}, \\ 
Razine Moundir Ghorab\textsuperscript{1}, Miquel Moreto\textsuperscript{1}\textsuperscript{2}, Gokcen Kestor\textsuperscript{1}, Dario Garcia-Gasulla\textsuperscript{1}}
\IEEEauthorblockA{\{arnau.ayguade, miquel.alberti, cristian.gutierrez, emanuele.parisi, \\
moundir.ghorab, miquel.moreto, gokcen.kestor, dario.garcia\}@bsc.es}
\IEEEauthorblockA{\textsuperscript{1}\textit{Barcelona Supercomputing Center, } \textsuperscript{2}\textit{Universitat Politècnica de Catalunya}\\
}

}

\maketitle

\begin{abstract}
Large Language Models (LLMs) have rapidly improved in performance across code-related tasks, making their integration into Register Transfer Level (RTL) development increasingly attractive. Mimicking the behavior of inline code assistants, many benchmarks evaluate LLMs' capabilities in code completion, either assessing the generation of entire hardware modules or the completion of a single line within a module. However both of these approaches lack the ability to control the granularity of the code-completion sample size and the syntactic range of completions. To overcome these limitations, we present a framework for language-agnostic rule completion (RuC), a grammar-driven, rule-selectable benchmark generator that automatically produces RTL code-completion tasks from a set of input hardware description sources. RuC uses the target Hardware Description Language (HDL) grammar to mask syntactically defined code regions and prompts a model to regenerate them using the surrounding unmasked code as context, enabling a controlled and scalable evaluation of the domain-specific model's code-understanding capabilities, ranging from assignments to the reconstruction of entire logic blocks. We use RuC to generate two SystemVerilog rule-completion benchmarks from the Tiny Tapeout shuttle \texttt{TT07} and the CVE2 RISC-V core to demonstrate RuC's applicability to a broad range of designs, and conduct a comparative study of the code completion capabilities of modern open-source LLMs across diverse settings. Results indicate that completion performance strongly depends on the model type, the grammatical structure of the masked region, and the prompting strategy. Specifically, the highest scores are obtained with Fill-in-the-Middle (FIM) prompting. These findings highlight the value of grammar-driven, arbitrarily granular benchmarks for meaningful evaluation of LLM capabilities in RTL development workflows.
\end{abstract}


\section{Introduction}

Large language models (LLMs) have demonstrated strong capabilities across a broad range of code-related tasks, including code generation from specification and context-aware completion~\cite{jiang2026survey}. This progress has sparked increased interest in employing LLMs as coding assistants in the electronic design automation (EDA) domain, and the community has introduced a variety of benchmarks that evaluate models’ capabilities in code understanding and generation under different settings~\cite{pan2025survey,he2025large,jha2025large,xu2025large}. Among these, code completion is particularly important, as it is the canonical setting for assessing whether a model can leverage contextual information to produce semantically correct code and directly mirrors the interactive use of LLMs as copilots during development. In this paradigm, a region of code is masked, and an LLM is tasked with reconstructing it using only the surrounding logic and overall project structure as contextual information~\cite{allam2024rtlrepo,ghorab2025notsotiny}. 

Existing code-completion benchmarks have converged into two principal families: Module Completion (MC) and Single-Line Completion (SLC)~\cite{liu2023verilogeval,allam2024rtlrepo,pinckney2025comprehensive,ghorab2025notsotiny,garciagasulla2025turtle}. In MC benchmarks, the complete implementation of a module within a design is masked, and the model must regenerate the entire module body by inferring its intended functionality from the surrounding design context. The objective is to test the model's global code-understanding capabilities by assessing whether it can reconstruct a module’s behavior from its interface and its integration into the broader project~\cite{ghorab2025notsotiny}. In contrast, SLC randomly selects a line within a target hardware module, masks the selected line together with the subsequent lines, and prompts the LLM to predict the next line given the available in-file prefix and additional repository-level context. This formulation is designed to evaluate the model’s understanding of local behavior and to approximate the incremental workflow typical of copilot-style assistance during RTL development~\cite{allam2024rtlrepo}. Despite their widespread adoption, both MC and SLC exhibit limitations. MC does not scale to complex designs because, in realistic hardware systems, a module implementation may span hundreds of lines of RTL code and include intricate control logic, making it impossible to reconstruct such a large region without any specification beyond contextual cues. SLC, on the other hand, relies on the notion of a "line," a rather arbitrary formatting unit rather than a formally defined HDL construct. Consequently, performance on SLC does not directly reveal the model’s ability to reproduce specific language features, as a randomly selected line may correspond to a trivial or semantically uninformative fragment. Additionally, the masking of all following context may make SLC impossible in some instances. Existing benchmarks force an \textit{all-or-nothing} choice between regenerating an entire module body or predicting a single line, and they fail to offer a nuanced and granular framework for assessing an LLM’s understanding of a project’s internal workings across semantically meaningful intermediate regions.

To address these limitations, we present RuC (Rule-based Completion), a grammar-driven framework that generates code-completion benchmarks by parsing input HDL descriptions and masking specific grammar rules, such as port declarations, assignments, or procedural blocks. RuC enables controllable difficulty through rule selection and masked-region size, supports different prompting strategies, and provides rigorous evaluation through compilation-based syntax checking (STX) and equivalence-oriented functional validation (EQV). In summary, rule completion enables:
\begin{itemize}
    \item \textbf{Difficulty Tuning:} The benchmark difficulty can be arbitrarily tuned by the choice of the grammar rules and region size candidate for masking.
    \item \textbf{Domain-Specific Evaluation:} Selectively evaluate the model understanding of domain-relevant capabilities (e.g., datapath generation, module instantiation patterns) that cannot be isolated when the masked region is a random line or a full module implementation.
    \item \textbf{HDL-Agnostic:} This approach can be applied to every language for which a grammar definition is available.
\end{itemize}
To demonstrate the flexibility of this approach to a broad range of designs, we use RuC to derive two SystemVerilog rule-completion benchmarks from the Tiny Tapeout shuttle \texttt{TT07}~\cite{tinytapeout} and the CVE2 RISC-V core~\cite{schiavone2017slow}. Finally, we evaluate a selection of modern open-source LLMs on the generated benchmarks, highlighting how rule-completion performance varies with model size, prompting strategies, and the types of grammar rules selected. Results show that Fill-In-the-Middle (FIM) prompting improves performance. Moreover, the high variance in results across distinct grammatical rules underscores the value of our granular per-rule assessment, with \texttt{case\_statement} and \texttt{always\_construct} being the most challenging rules.

\section{The RuC Framework}
\label{sec:The RuC Framework}

This section describes how the RuC framework generates grammar-driven rule-completion benchmarks. Section~\ref{subsec:Grammars and Rule-Based Code Completion} introduces the concept of grammars and provides a practical example demonstrating how rule-completion samples are generated. Section~\ref{subsec:Task construction pipeline} details the core components of the RuC framework. Sections~\ref{subsec:Prompt construction} and~\ref{subsec:Verification pipeline} explain how RuC constructs prompts and checks the correctness of the generated code.

\subsection{Grammars and Rule-Based Code Completion}
\label{subsec:Grammars and Rule-Based Code Completion}

\begin{figure}[t]
    \centering
    \includegraphics[draft=false,width=0.65\columnwidth]{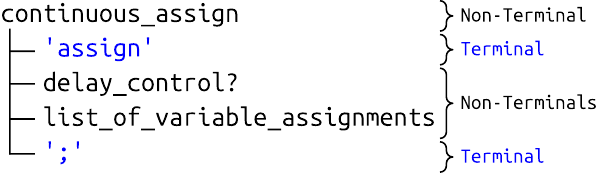}
    \caption{SystemVerilog rule for the continuous assignment non-terminal.}
    \label{fig:ContinuousAssign_Production}
\end{figure}

\begin{figure}[t]
    \centering
    \includegraphics[draft=false,width=0.55\columnwidth]{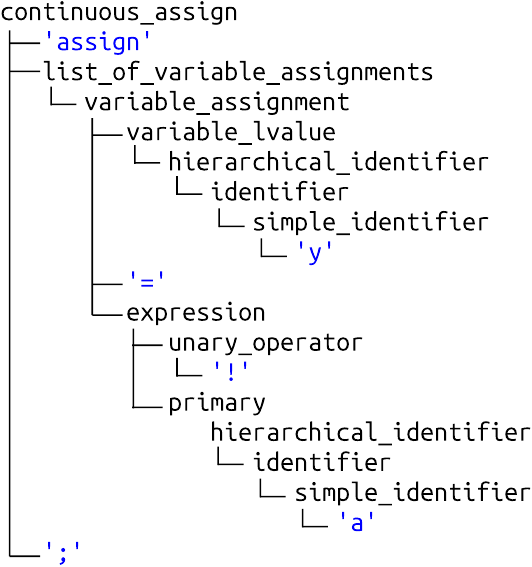}
    \caption{Parse tree obtained by parsing the SystemVerilog continuous assignment \texttt{assign y = !a;}. Terminals are highlighted in blue.}
    \label{fig:ContinuousAssign_ParseTree}
\end{figure}

In formal language theory, the syntax of programming and hardware description languages is defined by a finite set of production rules specified by a context-free grammar. A grammar is typically represented as a 4-tuple $G = (V, \Sigma, P, S)$, where $\Sigma$ denotes a finite set of terminal symbols, $V$ a finite set of non-terminal symbols, $P$ a finite set of productions, and $S \in V$ the start symbol. Terminals correspond to the lexemes appearing in valid sentences of the language; in RTL code, these include keywords, operators, and identifiers. Non-terminals represent abstract syntactic categories, each defining a set of strings and structuring the definition of valid constructs. The start symbol designates the language being defined, while the other non-terminals introduce auxiliary string classes used in its recursive specification. Each production, or rule, in $P$ consists of a head (the non-terminal being defined) and a body comprising a sequence of terminals and non-terminals. It specifies one admissible expansion of the head by leaving terminals unchanged and recursively substituting non-terminals with strings from their respective languages~\cite{hopcroft2012introduction}. As an illustrative example, consider the SystemVerilog continuous assignment \texttt{assign y = !a;}. Figure~\ref{fig:ContinuousAssign_Production} specifies the structure that every valid continuous assignment must follow~\footnote{According to the SystemVerilog grammar provided by ANTLR at \url{https://github.com/antlr/grammars-v4} (commit \texttt{962e91ce6c}).}. The non-terminal \texttt{continuous\_assign} is defined as a sequence consisting of the terminal \texttt{assign} keyword, an optional \texttt{delay\_control} non-terminal, the mandatory \texttt{list\_of\_variable\_assignments} non-terminal, and the terminating \texttt{;} terminal. Each non-terminal node is recursively expanded until only terminals appear at the leaves of the tree, and a depth-first traversal of the parse tree produces exactly the sequence of terminals composing the original input statement. The RuC framework leverages the parse tree to generate rule-completion benchmarks, shown in Figure~\ref{fig:ContinuousAssign_ParseTree}. RuC masks arbitrary portions of input code corresponding to user-selected grammar productions (i.e., sub-trees in the parse tree). Table~\ref{tab:continuous_assign} clarifies how rule-completion samples look, and how masking different grammar rules yields completion tasks of varying granularity. This grammar-driven approach enables fine-grained control over benchmark difficulty, allowing the complexity of the code-completion task to be systematically adjusted by selecting the grammar rule to mask.

\begin{table}[t]
    \centering
    \caption{Examples of increasingly difficult code-completion samples generated by RuC for different masked grammar rules.}
    \label{tab:continuous_assign}

    \begin{tabular}{@{}ll@{}}
        \toprule

        \multicolumn{1}{c}{Masked Rule}            &
        \multicolumn{1}{c}{Rule-Completion Sample} \\

        \midrule
        
        --                            & \texttt{assign y = !a;}      \\
        \texttt{simple\_identifier}   & \texttt{assign <MASK> = !a;} \\
        \texttt{expression}           & \texttt{assign y = <MASK>;}  \\
        \texttt{variable\_assignment} & \texttt{assign <MASK>;}      \\
        \texttt{continuous\_assign}   & \texttt{<MASK>}              \\

        \bottomrule

    \end{tabular}
\end{table}

\subsection{Task construction pipeline}
\label{subsec:Task construction pipeline}

\begin{figure*}[t]
    \centering
    \includegraphics[draft=false,width=0.9\textwidth]{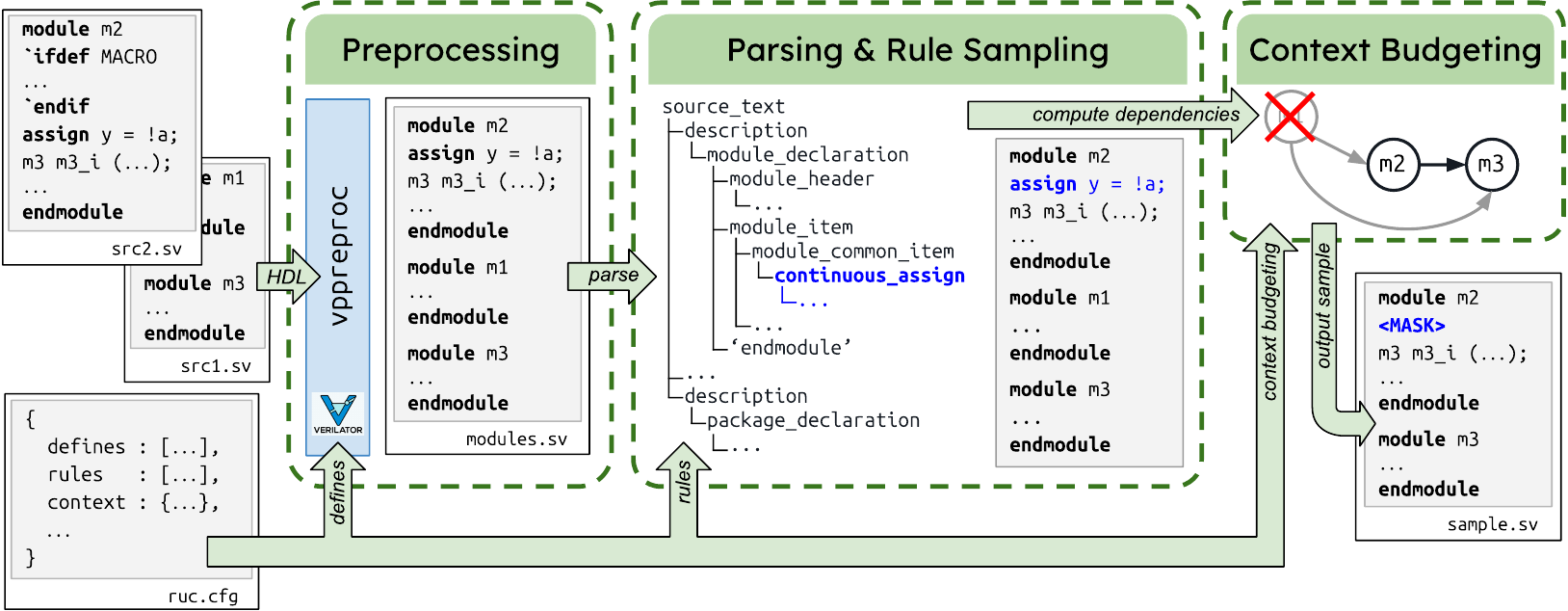}
    \caption{Overview of the task construction pipeline of the RuC framework for SystemVerilog sources. RuC generates rule-completion samples from a set of input sources and a pipeline configuration file through three stages: preprocessing, parsing and rule sampling, and context budgeting.}
    \label{fig:RuC_Framework}
\end{figure*}

The RuC framework implements an end-to-end processing pipeline to generate rule-completion samples from a set of HDL sources by exploiting the grammatical structure of the target hardware description language. An overview of the rule-completion sample generation pipeline is shown in Figure~\ref{fig:RuC_Framework}. RuC receives as input the HDL sources from which samples will be extracted, the metadata required to build the design (e.g., compiler definitions, include directories), and the set of grammatical rules that are candidates for masking. Based on these inputs, RuC parses the HDL sources and generates rule-completion samples for LLM evaluation. The RuC task construction pipeline comprises three main stages: preprocessing, parsing and rule sampling, and context budgeting. 

\subsubsection{Preprocessing}

During the preprocessing stage, the list of input sources, compiler definitions, and include directories is passed to a language preprocessor that resolves compiler directives and merges the sources into a single file. This step simplifies generating rule-completion samples from large codebases composed of multiple files scattered across different directories. It also facilitates prompting LLMs by providing a unified context that includes all design elements.

\subsubsection{Parsing \& Rule Sampling}

The parsing and rule sampling stage constitutes the core of the RuC framework. The source generated during preprocessing is first parsed to produce the corresponding parse tree. The framework then identifies and records the positions and occurrences of all user-selected grammatical rules that can serve as masking candidates. Rule-completion samples are generated by sampling and masking a specified number of occurrences for each selected rule, producing a set of rule-completion tasks. Before proceeding further, the generated samples are passed through the formal verification pipeline described in Section~\ref{subsec:Verification pipeline} to ensure that the masking process does not select regions derived from dead code or constructs that are removed during elaboration, such as branches of inactive \texttt{if-generate} statements.

\subsubsection{Context Budgeting}

The context budgeting stage addresses the challenge of the limited context window in current LLMs when dealing with large hardware designs. RuC implements a two-step context reduction strategy. First, module dependencies are extracted from the parse tree of the preprocessed design. A module is considered dependent on any module that it instantiates, as well as on any package it imports. Next, the framework identifies the module or package containing each generated rule-completion sample and prunes from the sample context all modules and packages that are not dependencies of that module or package. The user may choose whether to retain only direct dependencies or recursively include dependencies of dependencies, trading context size for additional structural information. The final output of the RuC task construction pipeline consists of a set of masked sources and the original preprocessed reference source, which serves as the ground truth during evaluation.

\subsection{Prompt construction}
\label{subsec:Prompt construction}

\begin{figure}[t]
    \centering
    \includegraphics[draft=false,width=0.8\columnwidth]{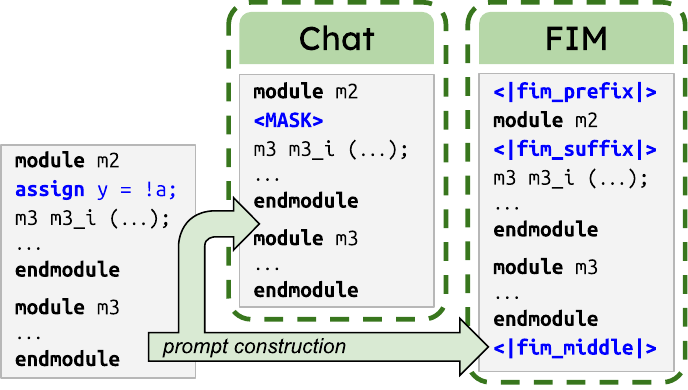}
    \caption{Overview of RuC prompt construction. After task construction, candidate samples are converted into either a chat-based or FIM-based prompt.}
    \label{fig:RuC_Prompting}
\end{figure}

Rule-completion tasks generated by the RuC framework fall within the class of code-infilling problems, where a model generates code at a specific location using both preceding and following context~\cite{fried2023incoder}. These tasks can be formulated in various ways depending on how the target model processes contextual information during training. To ensure fair evaluation aligned with the capabilities acquired during pretraining and fine-tuning, RuC supports multiple prompting paradigms that reflect common training strategies employed in modern code-oriented LLMs, as visualized in Figure~\ref{fig:RuC_Prompting}. The simplest approach is chat-based prompting. In this method, the selected grammar-rule region is removed from the source code and replaced with a placeholder token, such as \texttt{<MASK>}. The prompt then provides the model with the surrounding context and an instruction specifying that the generated code must integrate correctly with the existing implementation. This approach leverages the instruction-following capabilities of models fine-tuned on supervised instruction–response pairs. RuC additionally supports Fill-in-the-Middle (FIM) prompting, a recently adopted technique that explicitly trains models for infilling tasks by introducing dedicated FIM tokens~\cite{bavarian2022efficient}. In FIM prompting, the original source code is divided into three segments: the prefix representing the context preceding the middle region, the masked middle region corresponding to the selected grammar rule, and the suffix representing the remaining context. 
The prompt is constructed by concatenating the prefix and suffix segments and concluding with the middle token, after which the model generates the missing middle segment conditioned on the surrounding context.

\subsection{Verification pipeline}
\label{subsec:Verification pipeline}

\begin{figure}[t]
    \centering
    \includegraphics[draft=false,width=0.9\columnwidth]{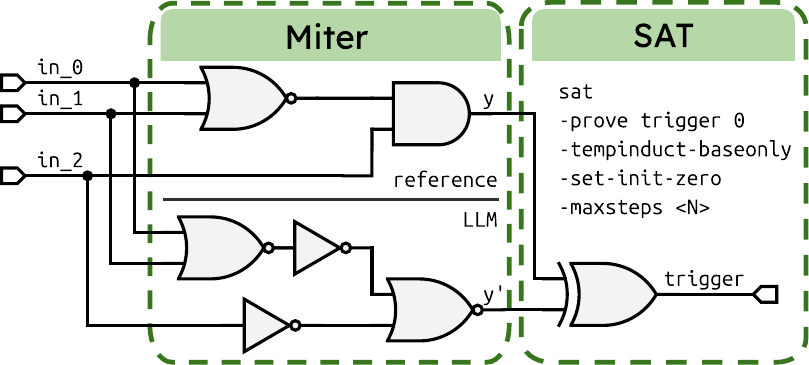}
    \caption{Overview of the RuC formal verification pipeline. A SAT solver is used to prove that the two circuits cannot produce different outputs.}
    \label{fig:RuC_Verification}
\end{figure}

After an LLM generates a candidate completion for a dataset sample, the RuC framework evaluates its syntactic and functional correctness. Syntactic validity is assessed by reinserting the generated snippet into the original source file and linting the design. This process ensures that the completion adheres to language rules and integrates consistently with the surrounding context. To evaluate functional correctness without relying on testbenches, which may vary in availability and quality across projects, the RuC framework employs formal verification techniques to establish functional equivalence between the LLM-generated completion and the original implementation. Specifically, RuC constructs a miter circuit from both the original and generated designs by providing identical inputs and comparing their outputs as illustrated in Figure~\ref{fig:RuC_Verification}. Whenever the \texttt{trigger} signal rises to 1, it indicates that an input sequence has caused the two circuits to produce different outputs. A satisfiability solver (SAT) is then used to prove that the trigger signal remains zero, thereby demonstrating that no input sequence exists that can cause a behavioral mismatch between the two designs. In practice, RuC tries to verify equivalence by temporal induction but evaluates only the base case of the proof, ignoring the induction step~\cite{een2003temporal}. This approach avoids false negatives in large designs, where SAT may fail to prove the induction step within the number of timesteps specified by the user.

\section{Experimental Results}
\label{sec:Experimental Results}

This section presents the evaluation of a set of LLMs with different prompting strategies on rule-completion benchmarks generated by RuC. While we rely on SystemVerilog, RuC is completely general, and projects in any HDL can be used for building rule-completion benchmarks. Section~\ref{subsec:Benchmark characterization and harness description} introduces the grammar rules used to construct the benchmarks, and it characterizes the considered codebases. Section~\ref{subsec:Prompting strategy and model type ablation study} studies the effects of prompting strategies, while Section~\ref{subsec:Grammar rule performance analysis} evaluates the top-performing models across different grammar rules, providing a detailed rule-level performance analysis.

\subsection{Benchmark characterization and harness description}
\label{subsec:Benchmark characterization and harness description}

We focus our evaluation on SystemVerilog constructs commonly found in the behavioural description of RTL modules, as summarized in Table~\ref{tab:Rules_SVType}. The \texttt{PORT} and \texttt{PARAM} rules assess LLMs' ability to reconstruct missing elements of module interfaces. The \texttt{INST} rule addresses module instantiations and evaluates whether models accurately reproduce logic-reuse patterns. The remaining constructs, \texttt{CONT}, \texttt{BLK}, \texttt{NBLK}, \texttt{COND}, \texttt{CASE}, and \texttt{ALWS}, test the reconstruction of behavioral logic, ranging from simple assignment to conditional and case-based control structures, culminating in complete \texttt{always} blocks. To construct rule-completion tasks, we select two codebases: the collection of designs from the Tiny Tapeout shuttle \texttt{TT07}~\cite{tinytapeout} and the CVE2 RISC-V core~\cite{schiavone2017slow}. Tiny Tapeout is a collaborative initiative that allows designers to submit open-source digital circuits for fabrication via periodic shuttles. We focus on \texttt{TT07} because it contains the highest number of occurrences of the selected grammatical constructs among available shuttles. CVE2 is an industry-grade RISC-V core maintained by OpenHWGroup. In the Tiny Tapeout and CVE2 projects, we employ RuC's context budgeting functionality to generate samples with up to 32\,000 tokens of context. Moreover, for CVE2, we exclude samples with a token size below 4\,000 to avoid evaluating the rule completion of small modules with limited functionality. The rule occurrence counts, and average token and line counts per rule of the resulting benchmarks are summarized in Table~\ref{tab:Rules_Statistics}. We configure the RuC pipeline to select at most 100 occurrences per rule. We employ \texttt{vppreproc} for source preprocessing and ANTLR v4.13 as parser generator~\cite{parr2011ll}. Our pipeline checks syntax correctness by linting the LLM-generated design with Verilator and verifies functionality by loading the design into Yosys~\cite{wolf2013yosys} and checking equivalence with its built-in SAT solver. Finally, we evaluate the rule-completion performance of five open-source LLMs, considering three compact, coding-oriented models suitable for local copilot-style deployment: Qwen2.5 Coder 14B, Seed Coder 8B, and Qwen3 Coder 30B A3B, and two larger, state-of-the-art models: Qwen3 Coder 480B A35B and DeepSeek v3.1 Terminus.

\begin{table}[t]
    \centering
    \caption{List of considered SystemVerilog grammar rules.}
    \label{tab:Rules_SVType}

    \begin{tabular}{@{}ll@{}}
        \toprule

        \multicolumn{1}{c}{Grammar Rule Name} &
        \multicolumn{1}{c}{Acronym} \\

        \midrule

        \texttt{ansi\_port\_declaration}                   & \texttt{PORT}  \\
        \texttt{parameter\_declaration}                    & \texttt{PARAM} \\
        \texttt{module\_program\_interface\_instantiation} & \texttt{INST}  \\
        \texttt{continous\_assignment}                     & \texttt{CONT}  \\
        \texttt{nonblocking\_assignment}                   & \texttt{NBLK}  \\
        \texttt{blocking\_assignment}                      & \texttt{BLK}   \\
        \texttt{conditional\_statement}                    & \texttt{COND}  \\
        \texttt{case\_statement}                           & \texttt{CASE}  \\
        \texttt{always\_construct}                         & \texttt{ALWS}  \\

        \bottomrule

    \end{tabular}
\end{table}

\begin{table}[t]
    \centering
    \caption{Grammar Rules Frequency and Average Length and Tokens for TinyTapeout and CVE2 Benchmarks}
    \label{tab:Rules_Statistics}

    \begin{tabular}{@{}lrrr rrr@{}}
        \toprule
        
        \multicolumn{1}{c}{\multirow{2}{*}{\vspace{-.05in}Rules}} &
        \multicolumn{3}{c}{Tiny Tapeout~\cite{tinytapeout}}        & 
        \multicolumn{3}{c}{CVE2~\cite{schiavone2017slow}}         \\
        
        \cmidrule(lr){2-4}
        \cmidrule(lr){5-7}
        
              &
        Count & LoC & Tokens &
        Count & LoC & Tokens \\

        \midrule

        \texttt{PORT}  & 1218 &  1.0 &   5.5 & 408 &  1.0 &   9.8 \\
        \texttt{PARAM} &  139 &  1.1 &   8.3 & 389 &  1.0 &  13.0 \\
        \texttt{INST}  &  360 &  6.0 &  56.3 &  27 & 12.6 & 103.0 \\
        \texttt{CONT}  &  790 &  1.2 &  25.9 & 524 &  1.3 &  22.5 \\
        \texttt{NBLK}  & 1982 &  1.0 &   8.3 & 134 &  1.0 &   8.7 \\
        \texttt{BLK}   &  536 &  1.0 &  10.2 &1331 &  1.1 &  14.8 \\
        \texttt{COND}  &  508 & 13.9 & 135.5 & 264 &  9.2 & 121.2 \\
        \texttt{CASE}  &   56 & 40.9 & 364.1 &  85 & 50.5 & 701.7 \\
        \texttt{ALWS}  &  164 & 30.9 & 283.1 &  95 & 39.5 & 501.0 \\

        \bottomrule
    \end{tabular}
\end{table}

\subsection{Prompting strategy and model type ablation study}
\label{subsec:Prompting strategy and model type ablation study}

\begin{table}[t]
    \centering
    \caption{Average Syntax (STX) and Functionality (EQV) Scores for the Tiny Tapeout Benchmark. Best configuration for each model in bold.}
    \label{tab:avg_scores}

    \begin{tabular}{lrr}
        \toprule

        \multicolumn{1}{c}{\multirow{2}{*}{\vspace{-.05in}Model}} &
        \multicolumn{2}{c}{Tiny Tapeout~\cite{tinytapeout}} \\

        \cmidrule(lr){2-3}

        &
        \multicolumn{1}{c}{STX} & \multicolumn{1}{c}{EQV} \\
        
        \midrule

        \textbf{Qwen2.5 Coder 14B Base FIM}          & \textbf{90.1\%} & \textbf{60.2\%} \\
                Qwen2.5 Coder 14B Instruct FIM       &         61.2\%  &         49.2\%  \\
                Qwen2.5 Coder 14B Instruct CHAT      &         79.3\%  &         58.9\%  \\
        \midrule
        \textbf{Seed Coder 8B Base FIM}              & \textbf{74.7\%} & \textbf{58.6\%} \\
                Seed Coder 8B Instruct FIM           &         46.3\%  &         36.3\%  \\
                Seed Coder 8B Instruct CHAT          &         38.3\%  &         23.6\%  \\
        \midrule
        \textbf{DeepSeek V3.1 Terminus Instruct FIM} & \textbf{93.5\%} & \textbf{74.0\%} \\
                DeepSeek V3.1 Terminus Instruct CHAT &         92.2\%  &         73.9\%  \\
        \midrule
        \textbf{Qwen3 Coder 30B A3B Instruct FIM}    & \textbf{83.0\%} & \textbf{48.3\%} \\
                Qwen3 Coder 30B A3B Instruct CHAT    &         65.5\%  &         46.5\%  \\
        \midrule
        \textbf{Qwen3 Coder 480B A35B Instruct FIM}  & \textbf{98.4\%} & \textbf{76.7\%} \\
                Qwen3 Coder 480B A35B Instruct CHAT  &         91.7\%  &         71.0\%  \\

        \bottomrule
    \end{tabular}
\end{table}

As part of the first experimental campaign, we conducted a thorough assessment to study the influence of the model and prompt type under the same benchmark. We distinguished between base models, pre-trained on unstructured corpora, and their instruction-tuned variants, which are fine-tuned to follow natural language commands, and evaluated them under two prompting paradigms: FIM-based and Chat-based. We excluded the combination of base models with Chat-based templates from the evaluation because base models are not trained to follow instructions. The results can be seen in Table~\ref{tab:avg_scores}, with the best performing variants for each model in bold. We employed the specialized FIM tokens and ordering specific to each model. The data suggest that FIM-based prompting is the most effective approach, as the task format aligns with the FIM training objective that these models were optimized for during pre-training~\cite{bavarian2022efficient}. When comparing base and instruction-tuned variants of the same model, the base model achieves superior performance, highlighting the prevalence of next-token prediction in the rule-completion task. Existing instruction-tuning techniques, which improve adherence to commands in general code generation, can degrade FIM performance, forcing a trade-off between instruction-following and infilling capabilities~\cite{FIMParadigm}.

\subsection{Grammar rule performance analysis}
\label{subsec:Grammar rule performance analysis}

\begin{figure*}[t]
    \centering
    \includegraphics[draft=false,width=0.95\textwidth]{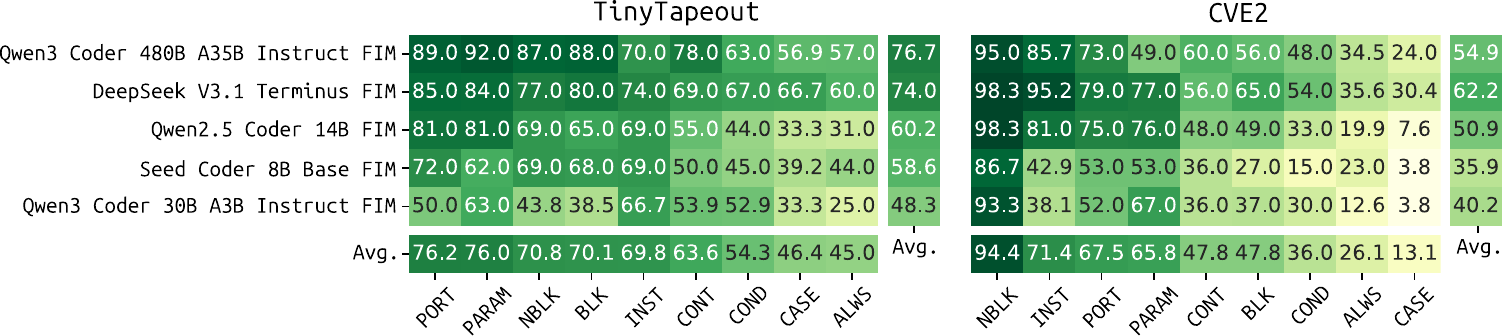}
    \caption{Rule-completion performance across models and grammar rules. Each cell reports the percentage of tasks for which the generated implementation is formally equivalent to the reference. Rules are ordered by their average EQV success rate across models, which is reported in the bottom panel.}
    \label{fig:Results_Heatmap}
\end{figure*}

This section presents the functional correctness of rule completions generated by LLMs. Each row of Figure~\ref{fig:Results_Heatmap} summarizes the best-performing model variant and prompting strategy for each of the five evaluated model families. Across the evaluated models, the average performance difference between the easiest and most challenging rules reaches 31.2\% in the \texttt{TT07} benchmark and 81.3\% in the CVE2 benchmark. This substantial variability underscores the utility of grammar-driven rule-completion benchmarks for designing evaluation tasks with adjustable difficulty, thereby enabling systematic scaling of benchmark complexity. A clear trend is observed between the size of the masked region and the likelihood that a model correctly reconstructs it. The easiest rules correspond to \texttt{PORT} and \texttt{PARAM}, whose correct reconstruction can often be inferred from surrounding module usage and previously defined interfaces. Assignment-related rules, such as \texttt{CONT}, \texttt{BLK}, and \texttt{NBLK}, are moderately more challenging. Although these constructs typically span a single line, they require the model to infer behavioral logic from the surrounding module implementation. The most difficult rules are \texttt{COND}, \texttt{CASE}, and \texttt{ALWAYS}, which generally correspond to larger code regions and require reconstructing more complex behavioral logic. An anomalous behavior is observed for the \texttt{NBLK} rule in the CVE2 benchmark, where performance is significantly higher than expected. This phenomenon can be attributed to the fact that most sampled non-blocking assignments implement register-update logic following highly regular patterns, such as \texttt{<signal>\_q <= <signal>\_d}. This behavior highlights that rule-completion accuracy is influenced not only by the grammatical rule under evaluation but also by stylistic regularities within the target codebase. Finally, comparisons across model families reveal that larger models achieve higher overall accuracy. However, in certain cases, this rule does not hold. In the \texttt{TT07} benchmark, Seed Coder outperforms Qwen3 30B and achieves a performance comparable to Qwen2.5 despite being smaller than both. These results suggest that model size alone does not fully account for performance differences in rule-completion tasks, thereby opening the door to rule-specific LLM selection.

\section{Related Works}
\label{sec:Related Works}

Initial efforts to benchmark LLMs in the domain of RTL used highly curated datasets to produce high-quality evaluation samples. However, the featured designs were simple, and intensive human intervention was required to construct and verify these datasets, which severely limited the scalability of such approaches and constrained their ability to evaluate today's more advanced models at scale~\cite{liu2023verilogeval,thakur2024verigen,lu2024rtllm}.

Recently, several RTL code-completion benchmarks have been introduced to evaluate LLMs’ ability to generate code from the surrounding hardware context. RTL-Repo~\cite{allam2024rtlrepo} evaluates LLMs at single-line completion tasks built from large-scale Verilog repositories. Given a code source, RTL-Repo samples non-empty, non-comment lines from different files as prediction targets, providing the model with the full repository context and the current file's HDL up to the selected line. While this approach allows benchmarking scalability across complex codebases, the random extraction of target lines may yield ill-posed or unsolvable tasks, as the code following the target line is discarded without regard to downstream dependencies. Additionally, the absence of a dedicated prompting strategy may hinder LLMs' correct interpretation of the task’s goal. Finally, the reliance on exact-match and edit-similarity is also limiting in the RTL domain, as these metrics do not award partial credit for syntactically correct outputs and do not rigorously evaluate functional correctness, potentially leading to false negatives~\cite{CodeScore}. In contrast, the proposed RuC framework preserves the scalability advantages of RTL-Repo while addressing these limitations through grammar-driven target selection, the evaluation of the prompting strategy, and the consideration of both syntactic and functional correctness using appropriate hardware verification tools.

NotSoTiny~\cite{ghorab2025notsotiny} is a large-scale, ``living'' benchmark for RTL code completion that mitigates limitations in current hardware datasets, such as small scale, shallow verification, and data contamination. It features 1{,}114 deduplicated tasks derived from real-world Tiny Tapeout designs, and it centers on Module Completion, requiring LLMs to reconstruct missing modules by inferring their behavior solely from the surrounding system implementation. To remain ``contamination-resilient'' against future LLM training data, the benchmark employs an automatic pipeline that updates the dataset with new Tiny Tapeout fabrication shuttles. Furthermore, the benchmark elevates verification standards by replacing limited simulation testbenches with rigorous formal verification.

The CVDP benchmark~\cite{pinckney2025comprehensive} introduces a robust dataset of 783 complex problems spanning 13 RTL-related tasks, all manually crafted by a large team of experienced hardware engineers. Its primary focus is on agentic evaluation, creating environments where AI agents can inspect mini-repositories and invoke external tools. It also includes tasks for LLMs in a single-turn setting, 94 of which target code completion.

\section{Conclusions}
\label{sec:Conclusions}

This work presents RuC, a framework for creating grammar-driven, rule-completion benchmarks, enabling fine-grained, scalable evaluations of LLMs' generation and understanding capabilities for RTL code. RuC features an HDL-agnostic task construction pipeline, a prompting strategy tailored for rule completion, and a robust verification pipeline that tests syntactic and functional correctness via equivalence checking. We demonstrate the framework's flexibility by creating two SystemVerilog benchmarks from open-source designs and running them across a variety of state-of-the-art LLMs. Results show substantial performance gaps across rules, underscoring the need for a grammar-driven evaluation, and reveal that infilling prompting is more effective than standard chat-based prompting. Moreover, this work provides a foundation for future research evaluating how LLMs understand meaningful blocks of HDL code across controllable context lengths, which is essential for developing copilot-style assistants for RTL development. “The RuC framework source code is available on GitHub\footnote{\githuburl}, and the datasets used in this work are available on Hugging Face\footnote{\huggingfaceurl}.”

\section{Acknowledgments}
\label{sec:Acknowledgments}

This work is supported by the AI4S fellowships awarded to Gokcen Kestor, Emanuele Parisi, Razine Moundir Ghorab, Cristian Gutierrez Gomez, and Miquel Albertí Binimelis under the “Generación D” initiative of Red.es and the Ministerio para la Transformación Digital y de la Función Pública for talent attraction under grant C005/24-ED CV1
funded by the European Union through the NextGenerationEU program and PRTR. 
It is also partially supported by the ELLIOT project funded by the European Union under grant agreement No. 101214398, and by project PID2023-146511NBI00 funded by the Spanish Ministry of Science, Innovation and Universities MCIU /AEI /10.13039/501100011033, and by the EU ERDF. 

Finally, we thank the Operations department at BSC for their technical support. We also acknowledge Bernat Homs and Serik Perez for their valuable discussions.

\bibliographystyle{IEEEtran}
\bibliography{main}

\end{document}